


\def\oneandahalfspace{\baselineskip=1.15\normalbaselineskip plus 1pt
\lineskip=2pt\lineskiplimit=1pt}

\def\nl{\hfil\break}

\def\nofirstpagenoten{\nopagenumbers\footline={\ifnum\pageno>1\tenrm
\hss\folio\hss\fi}}
\def\nofirstpagenotwelve{\nopagenumbers\footline={\ifnum\pageno>1\twelverm
\hss\folio\hss\fi}}
\def\leaderfill{\leaders\hbox to 1em{\hss.\hss}\hfill}
\def\ft#1#2{{\textstyle{{#1}\over{#2}}}}
\def\frac#1/#2{\leavevmode\kern.1em
\raise.5ex\hbox{\the\scriptfont0 #1}\kern-.1em/\kern-.15em
\lower.25ex\hbox{\the\scriptfont0 #2}}
\def\sfrac#1/#2{\leavevmode\kern.1em
\raise.5ex\hbox{\the\scriptscriptfont0 #1}\kern-.1em/\kern-.15em
\lower.25ex\hbox{\the\scriptscriptfont0 #2}}


\parindent=20pt
\def\narrow{\advance\leftskip by 40pt \advance\rightskip by 40pt}

\def\nonarrower{\advance\leftskip by -40pt\advance\rightskip by -40pt}

\def\boxit#1{\vbox{\hrule\hbox{\vrule\kern3pt
        \vbox{\kern3pt#1\kern3pt}\kern3pt\vrule}\hrule}}

\def\gtorder{\mathrel{\raise.3ex\hbox{$>$}\mkern-14mu
             \lower0.6ex\hbox{$\sim$}}}
\def\ltorder{\mathrel{\raise.3ex\hbox{$<$}|mkern-14mu
             \lower0.6ex\hbox{\sim$}}}
\def\dalemb#1#2{{\vbox{\hrule height .#2pt
        \hbox{\vrule width.#2pt height#1pt \kern#1pt
                \vrule width.#2pt}
        \hrule height.#2pt}}}

\font\fourteentt=cmtt10 scaled \magstep2
\font\fourteenbf=cmbx12 scaled \magstep1
\font\fourteenrm=cmr12 scaled \magstep1
\font\fourteeni=cmmi12 scaled \magstep1
\font\fourteenss=cmss12 scaled \magstep1
\font\fourteensy=cmsy10 scaled \magstep2
\font\fourteensl=cmsl12 scaled \magstep1
\font\fourteenex=cmex10 scaled \magstep2
\font\fourteenit=cmti12 scaled \magstep1
\font\twelvett=cmtt10 scaled \magstep1 \font\twelvebf=cmbx12
\font\twelverm=cmr12 \font\twelvei=cmmi12
\font\twelvess=cmss12 \font\twelvesy=cmsy10 scaled \magstep1
\font\twelvesl=cmsl12 \font\twelveex=cmex10 scaled \magstep1
\font\twelveit=cmti12
\font\tenss=cmss10
 
 \font\ninebf=cmbx7 scaled \magstep1
\font\ninerm=cmr7 scaled \magstep1 \font\ninei=cmmi7 scaled \magstep1
\font\ninesy=cmsy7 scaled \magstep1 
\font\eightrm=cmr7 scaled 1140 
 
\font\sevenbf=cmbx7 \font\sevenrm=cmr7 \font\seveni=cmmi7
\font\sevensy=cmsy7 

\catcode`@=11
\newskip\ttglue
\newfam\ssfam

\def\fourteenpoint{\def\rm{\fam0\fourteenrm}
\textfont0=\fourteenrm \scriptfont0=\tenrm \scriptscriptfont0=\sevenrm
\textfont1=\fourteeni \scriptfont1=\teni \scriptscriptfont1=\seveni
\textfont2=\fourteensy \scriptfont2=\tensy \scriptscriptfont2=\sevensy
\textfont3=\fourteenex \scriptfont3=\fourteenex \scriptscriptfont3=\fourteenex
\def\it{\fam\itfam\fourteenit} \textfont\itfam=\fourteenit
\def\sl{\fam\slfam\fourteensl} \textfont\slfam=\fourteensl
\def\bf{\fam\bffam\fourteenbf} \textfont\bffam=\fourteenbf
\scriptfont\bffam=\tenbf \scriptscriptfont\bffam=\sevenbf
\def\tt{\fam\ttfam\fourteentt} \textfont\ttfam=\fourteentt
\def\ss{\fam\ssfam\fourteenss} \textfont\ssfam=\fourteenss
\tt \ttglue=.5em plus .25em minus .15em
\normalbaselineskip=16pt
\abovedisplayskip=16pt plus 4pt minus 12pt
\belowdisplayskip=16pt plus 4pt minus 12pt
\abovedisplayshortskip=0pt plus 4pt
\belowdisplayshortskip=9pt plus 4pt minus 6pt
\parskip=5pt plus 1.5pt
\setbox\strutbox=\hbox{\vrule height12pt depth5pt width0pt}
\let\sc=\tenrm
\let\big=\fourteenbig \normalbaselines\rm}
\def\fourteenbig#1{{\hbox{$\left#1\vbox to12pt{}\right.\n@space$}}}

\def\twelvepoint{\def\rm{\fam0\twelverm}
\textfont0=\twelverm \scriptfont0=\ninerm \scriptscriptfont0=\sevenrm
\textfont1=\twelvei \scriptfont1=\ninei \scriptscriptfont1=\seveni
\textfont2=\twelvesy \scriptfont2=\ninesy \scriptscriptfont2=\sevensy
\textfont3=\twelveex \scriptfont3=\twelveex \scriptscriptfont3=\twelveex
\def\it{\fam\itfam\twelveit} \textfont\itfam=\twelveit
\def\sl{\fam\slfam\twelvesl} \textfont\slfam=\twelvesl
\def\bf{\fam\bffam\twelvebf} \textfont\bffam=\twelvebf
\scriptfont\bffam=\ninebf \scriptscriptfont\bffam=\sevenbf
\def\tt{\fam\ttfam\twelvett} \textfont\ttfam=\twelvett
\def\ss{\fam\ssfam\twelvess} \textfont\ssfam=\twelvess
\tt \ttglue=.5em plus .25em minus .15em
\normalbaselineskip=14pt
\abovedisplayskip=14pt plus 3pt minus 10pt
\belowdisplayskip=14pt plus 3pt minus 10pt
\abovedisplayshortskip=0pt plus 3pt
\belowdisplayshortskip=8pt plus 3pt minus 5pt
\parskip=3pt plus 1.5pt
\setbox\strutbox=\hbox{\vrule height10pt depth4pt width0pt}
\let\sc=\ninerm
\let\big=\twelvebig \normalbaselines\rm}
\def\twelvebig#1{{\hbox{$\left#1\vbox to10pt{}\right.\n@space$}}}

\def\tenpoint{\def\rm{\fam0\tenrm}
\textfont0=\tenrm \scriptfont0=\sevenrm \scriptscriptfont0=\fiverm
\textfont1=\teni \scriptfont1=\seveni \scriptscriptfont1=\fivei
\textfont2=\tensy \scriptfont2=\sevensy \scriptscriptfont2=\fivesy
\textfont3=\tenex \scriptfont3=\tenex \scriptscriptfont3=\tenex
\def\it{\fam\itfam\tenit} \textfont\itfam=\tenit
\def\sl{\fam\slfam\tensl} \textfont\slfam=\tensl
\def\bf{\fam\bffam\tenbf} \textfont\bffam=\tenbf
\scriptfont\bffam=\sevenbf \scriptscriptfont\bffam=\fivebf
\def\tt{\fam\ttfam\tentt} \textfont\ttfam=\tentt
\def\ss{\fam\ssfam\tenss} \textfont\ssfam=\tenss
\tt \ttglue=.5em plus .25em minus .15em
\normalbaselineskip=12pt
\abovedisplayskip=12pt plus 3pt minus 9pt
\belowdisplayskip=12pt plus 3pt minus 9pt
\abovedisplayshortskip=0pt plus 3pt
\belowdisplayshortskip=7pt plus 3pt minus 4pt
\parskip=0.0pt plus 1.0pt
\setbox\strutbox=\hbox{\vrule height8.5pt depth3.5pt width0pt}
\let\sc=\eightrm
\let\big=\tenbig \normalbaselines\rm}
\def\tenbig#1{{\hbox{$\left#1\vbox to8.5pt{}\right.\n@space$}}}
\let\rawfootnote=\footnote \def\footnote#1#2{{\rm\parskip=0pt\rawfootnote{#1}
{#2\hfill\vrule height 0pt depth 6pt width 0pt}}}

\def\ft#1#2{{\textstyle{{#1}\over{#2}}}}

\def\semidirprod{\rlap{\ss C}\raise1pt\hbox{$\mkern.75mu\times$}}

\def\for{\lower6pt\hbox{$\Big|$}}

\nofirstpagenotwelve
\twelvepoint
\def\fish{\kern-.25em{\phantom{abcde}\over \phantom{abcde}}\kern-.25em}

\def\b{\beta}

\def\p{\partial}

\def\12{\delta^2 (\xi -\xi')}
\def\23{\delta^2 (\xi' -\xi'')}
\def\31{\delta^2 (\xi'' -\xi)}

\def\half{{1\over 2}}
\oneandahalfspace


\overfullrule=0pt
\def\dg{$^\dagger$}

\line{\hfil CTP-TAMU-13/92}
\vskip 1.0truein

\centerline{\bf  Area-Preserving Diffeomorphisms, $w_\infty$ Algebras and
$w_\infty$ Gravity
{\footnote\dg {Lectures given at the Trieste Summer School in High Energy
Physics,
July 1991.}}}  \vskip .90truein

\centerline{ E. Sezgin {\footnote{$^*$} {Work supported in part by NSF grant
PHY-9106593}} }
\bigskip

\centerline{Center for Theoretical Physics}
\centerline{Physics Department}
\centerline{Texas A\&M University} \centerline{College Station, Texas 77843}
\vskip .4truein
\centerline{\bf  February 1991}
\vskip .5truein

\centerline{\bf ABSTRACT}
\medskip

The $w_\infty$ algebra is a particular generalization of the Virasoro
algebra with generators of higher spin $2,3,...,\infty$. It can be viewed as
the algebra of a class of functions,  relative
to a Poisson bracket, on a  suitably chosen surface. Thus, $w_\infty$ is a
special case of area-preserving diffeomorphisms of an arbitrary surface. We
review various aspects of area-preserving
diffeomorphisms, $w_\infty$ algebras and $w_\infty$ gravity. The topics
covered include a) the structure of the algebra of area-preserving
diffeomorphisms with central extensions and their relation to  $w_\infty$
algebras, b) various generalizations of $w_\infty$ algebras, c) the structure
of $w_\infty$ gravity and its geometrical aspects, d) nonlinear  realizations
of $w_\infty$ symmetry and e) various quantum realizations of $w_\infty$
symmetry.

\vfill\eject

\noindent {\bf 1. Introduction}
\bigskip
Area-preserving diffeomorphisms of a surface $\Sigma$, denoted by
${\rm SDiff}(\Sigma)$, arise in
diverse areas of theoretical and mathematical physics. For example, since
they are the canonical
transformations which preserve the Poisson bracket, they naturally arise in
theory of dynamical
systems [1]. In the context of high
energy physics, the area-preserving diffeomorphisms of a 2-sphere,
${\rm SDiff}(S^2)$, has been
encountered in the theory of relativistic membranes
\footnote {$^\dagger$}{The generalization of this
group for higher dimensional extended objects, which are known as the
$p$-branes, is the
volume-preserving diffeomorphisms of a $p$ dimensional manifold [3].}[2].
It arises as a subgroup of the diffeomorphisms of
the 3-dimensional world-volume in a light-cone gauge. It was
observed that (in the case of spherical and toroidal membranes  at least)
the structure
constants of
the area-preserving diffeomorphisms are those of $SU(N)$  in the
$N\rightarrow \infty$ limit
(in the case of a torus there are the additional global diffeomorphisms)
[2]. This fact was
used in
regularizing the quantum theory of the supermembrane, by considering the
$SU(N)$ theory
and then taking the $N\rightarrow \infty$ limit [4].

Area-preserving diffeomorphisms of a 2-torus, ${\rm SDiff}(T^2)$, has been
studied as the
analog of the Virasoro symmetry [5]. It turns out that the generators of
the Virasoro algebra
(with or without central charge) can be constructed as a linear combination
of infinitely many
${\rm SDiff}(T^2)$ generators [6]. Later it was realised that the Virasoro
algebra can be
embedded in the area-preserving diffeomorphisms of the 2-plane,
${\rm SDiff}(R^2)$, in a
manifest manner [7]. In
this case, all the other generators of the algebra form a representation of
the Virasoro
algebra and
thus can be viewed as generators with  conformal spins $3,4,...,\infty$.
This observation,
opened the
way for borrowing many of the techniques used in the study of Virasoro
algebra and the
2D conformal
field theories.

There is a great deal of arbitrariness in choosing basis functions in order
to exhibit the
structure constants of the SDiff algebras explicitly. The problem becomes
more acute when
one deals
with noncompact surfaces such as $R^2$ or a cylinder, in which case one
usually works with
basis
functions which diverge at infinity. Nonetheless, there is a particular
choice of basis
elements
appropriate for a surface of cylindrical topology [8], which, though
divergent at infinity, it
yields a specific set of structure constants. Although, this particular
algebra is
referred to as $w_{\infty}$ or $w_{1+\infty}$ algebra (in the latter case
a spin 1 generator is
included), SDiff algebras for other kinds of surfaces, or for the same
surface but
with a different
choice of basis elements are also referred to loosely as $w_\infty$
algebras. Often it is not
so clear which one of these algebras are isomorphic into each other,
especially in the case of noncompact surfaces, and when infinitely many
redefinitions of the
generators is required. At any rate, the terminology of $w_\infty$ will
be used to refer to a wide class of higher spin algebras in two dimensions
which contain
the Virasoro algebra, and are connected  to certain area-preserving
diffeomorphisms in a way
to be specified carefully case by case, if needed.

In $w_\infty$ algebras the commutator of a spin $s$ generator with a spin
$s'$ generator yields a
spin $(s+s'-2)$ generator, and a central extension can arise only in the
commutator of two spin 2
generators. Denoting the Betti number of the surface $\Sigma$ by $b_1$,
the algebra
${\rm SDiff}(\Sigma)$ admits $b_1$ independent central extensions [9].
An extension of
$w_\infty$ in which the commutator a  a spin $s$ generator with a spin
$s'$ generator yields all
spins between spin 2 and spin $(s+s'-2)$ arise was constructed explicitly
in Refs.~[8,10]. This is
referred to as $W_\infty$ algebra and it does admit central extensions in
all spin sectors.

A new field theoretic realization of $w_\infty$ was found by constructing
the so called
$w_\infty$ gravity action [11], which is the analog of the Polyakov's
bosonic string action,
incorporating now the coupling of all higher spin world-sheet fields to
matter fields.
Interestingly enough, when quantized, the $w_\infty$ symmetry of this model
deforms into
$W_\infty$
symmetry [12]. In fact, it has been shown that this is a rather general
phenomenon
which arises even in simpler quantum mechanical systems in one dimension
[13]. A geometric
understanding of $w_\infty$ gravity is still lacking, but interesting
attempts
have been made in
this direction [14]. The $w_\infty$ transformation rules for a scalar field
have also been
interpreted in the context of a nonlinear realization of $w_\infty$ [11,15],
which we
shall discuss later.

There are a number of other areas in which $w_\infty$ symmetry arises. For
example, a 2D sigma model based on an area-preserving
diffeomorphisms of a 2-surface turns out to have field equations that are
 closely
related to the
self-dual gravity equations in 4D [16]. This is rather interesting, because
it suggests a 2D
conformal field theory approach to an interesting 4D quantum gravity
problem. In fact,
self-dual
gravity equations have been shown to arise in string theories with local
$N=2$ supersymmetry on the
world-sheet , and a target spacetime of $(2,2)$  signature [17].
Furthermore,
it has been shown that
the loop group of $w_\infty$ arises as a symmetry group in the $N=2$
superstring
theory [17].  More
recently, the loop algebra
of area-preserving diffeomorphisms of a 2-plane has been discovered as the
algebra of an infinite set
of spin 1 primary fields constructed out of a matter field and the Liouville
field [18]. A
possible connection between the $c=1$ bosonic string theory where this
symmetry arises and the
$N=2$ superstring theory has been suggested [17]. Finally, let us mention
that the
area-preserving diffeomorphisms have also been encountered as symmetry
groups in the study of
higher spin  field theories in $2+1$ dimensions [18].

In this brief review, we shall first
describe the algebraic structure of ${\rm SDiff}(\Sigma)$ and $w_\infty$.
After we
summarize various
generalizations of these algebras, we shall turn to their classical field
theoretic
realizations. In
particular, we shall describe the (super) $w_\infty$ gravity, nonlinear
realizations and various quantum realizations of $w_{1+\infty}$ symmetry.

\bigskip
\noindent{\bf 2. The structure of area-preserving diffeomorphisms and
$w_\infty$ algebras}
\bigskip
In two dimensions the area-form is the same as a nondegenerate closed
2-form, i.e. a
symplectic 2-form. Hence, the area-preserving diffeomorphisms are the
same as the
symplectic
diffeomorphisms, namely those diffeomorphisms which leave the symplectic
form on the
surface
invariant. Unless stated otherwise we shall consider
{\it compact\ orientable} surfaces.
Denoting the
symplectic form by $\Omega=\Omega_{ab}d\sigma^a d\sigma^b$ where
 $\sigma^a,\ a=1,2$
are the
coordinates of the surface, it must satisfy the condition
${\cal L}_\xi \Omega=0$,
where $ {\cal
L}_\xi$ denotes the Lie derivative along a vector $\xi^a$. The most general
solution
to this
condition takes the form
$$
    \xi^a=\Omega^{ab}(\partial_b \Lambda +\omega_b) \eqno (2.1)
$$
where $\Lambda(\sigma)$ is an arbitrary function and $\omega_a$ is a
harmonic
1-form, i.e. it is curl-free but it can not be written as a derivative of a
scalar
globally. As is
well known, on a genus $g$ surface there are $2g$ independent such harmonic
1-forms.
Let us
expand $\omega_a =c_r\omega_a^{(r)},\ r=1,...,2g$ where $c_r$ are arbitrary
constants and
$\omega_a^{(r)}$, are harmonic 1-forms which we normalize as
$\int_{\Sigma} \omega^{(r)}
\wedge \omega^{(s)} =\delta_{rs}$. We can represent the generators of
the area-preserving diffeomorphisms as follows
$$
    L_\Lambda =\Omega^{ab}\partial_b \Lambda \partial_a,\quad\quad
 P_r=\Omega^{ab}\omega_b^{(r)} \partial_a,\ \ r=1,...,2g  \eqno (2.2)
$$
These generators obey the algebra
$$
\eqalignno{
   [L_{\Lambda_1},L_{\Lambda_2}] &=L_{\Lambda_{12}},&(2.3a)\cr
  [P_r,L_\Lambda] &=L_{\Lambda_r},&(2.3b)\cr
  [P_r,P_s] &=L_{\Lambda_{rs}},&(2.3c)\cr}
$$
where the parameters appearing on the right-hand sides are given by
$$
\eqalignno{
\Lambda_{12} &=\Omega^{ab}\partial_b \Lambda_1 \partial_a \Lambda_2 &(2.4a)
\cr
\Lambda_r &=\Omega^{ab} \omega_b^{(r)}\partial_a\Lambda &(2.4b)\cr
\Lambda_{rs} &=\Omega^{ab}\omega_b^{(r)} \omega_a^{(s)}  &(2.4c)}
$$
{}From (2.4a) it is clear that  ${\rm SDiff}(\Sigma)$ algebra
is the algebra of functions on $\Sigma$ relative to a Poisson bracket
defined by $\{A,B\}
:=\Omega^{ab}\partial_b A\partial_a B$.

Let us now consider the quantum version of the above algebra whose
generators we shall
denote by ${\hat L}_\Lambda$ and ${\hat P}_r$. The central extension of
such an algebra
was determined in  Ref.~[9]. The result takes the form
$$
\eqalignno{
[\hat L_{\Lambda_1},\hat L_{\Lambda_2}] &=\hat L_{\Lambda_{12}} +
\int_\Sigma d^2\sigma \alpha\Omega^{ab}\omega_b
(\Lambda_1 {\buildrel \leftrightarrow\over \partial_a}\Lambda_2)
&(2.5a)\cr
[\hat P_r,\hat L_\Lambda] &=\hat L_{\Lambda_r}
-2\int_\Sigma d^2\sigma \alpha\Omega^{ab} \omega_b \omega_a^{(r)} \Lambda
&(2.5b)\cr
[\hat P_r,\hat P_s] &=\hat L_{\Lambda_{rs}}+\int_\Sigma
d^2\sigma \beta \Omega^{ab}\omega_b^{(r)}\omega_a^{(s)}, &(2.5c)\cr}
$$
where $\alpha$ and $\beta$ are two arbitrary densities.

In practice, it is very useful to expand the generators of this algebra in
a suitable
basis and to
find the structure constants explicitly. The choice of basis depends on the
topology
and geometry of
the surface $\Sigma$ and on the type of functions we wish to expand on it.
These issues
arise in the
very rich and fascinating subject of harmonic analysis. An extensive
discussion of this
subject is
beyond the scope of this paper. We shall be content by giving few examples
 here. Let us
consider
first a flat two torus $T^2$ defined by a square lattice with
side $2\pi$. We shall take the symplectic structure to be the Levi-Civita
symbol $\Omega^{ab}=\epsilon^{ab}$ with $\epsilon^{12}=-\epsilon^{21}=1$.
It is natural
to choose
as basis functions $e^{i\underline n\cdot \underline \sigma}$ where
$\underline n=(n_1,n_2)$ and
$n_1, n_2$ are integers.  Note that this is a complete and orthonormal
basis, and
furnish a unitary
representation of the torus group $U(1)\times U(1)$. In terms of these
basis functions
we may expand
the parameters  $\Lambda(\sigma)$ as
$$
\Lambda(\sigma)=\sum_{\underline n}
\Lambda_{\underline n}  e^{i\underline n\cdot \underline \sigma},\eqno(2.6)
$$
The torus has genus $1$, so there are two independent
harmonic $1$-forms.  We can parametrize the most general harmonic $1$-form
as $\omega_a=(a_1,a_2)$, where the components $a_1$ and $a_2$ are arbitrary
constants. Defining the Fourier components of the ${\rm SDiff}(T^2)$
generators as
follows
$$
\hat L_\Lambda= \sum_{\underline n} \Lambda_{ \underline n}\hat
L_{\underline n},
\eqno(2.7)
$$
we find that (2.5)  reduces in this special case to [9]
$$
\eqalignno{
[\hat L_{\underline n},\hat L_{\underline m}] &=\underline n\times
\underline m \;\hat L_{\underline n +\underline m}+ \underline a\times
\underline n
\; \delta_{\underline n +\underline m,\underline 0}  &(2.8a)\cr
[\hat P_r,\hat L_{\underline n}] &=n_r \hat L_{\underline
n}+b_r\delta_{\underline n,0} &(2.8b)\cr
[\hat P_r,\hat P_s] &=c\epsilon_{rs},&(2.8c)\cr}
$$
where $\underline n\times \underline m\equiv \epsilon^{ab} n_a m_b$ and
$b_r, c$ are constants given by $\b_r=\epsilon_{rs}c_s\Lambda_0
\int d^2\sigma \alpha$ and
$c=\int d^2\sigma \beta$, and $r,s=1,2$ since $b_1=2$ for a torus. Note
that
$L_{\underline 0}$
does not occur in the commutation relations (2.8). However, it can be
introduced on
the right hand
side of (2.8c) as $\epsilon_{rs}L_{\underline 0}$. It would be a central
charge
commuting with all
the generators. Moreover, the central extension on the right hand side of
(2.8c)
could be absorbed
in a redefinition of $L_{\underline 0}$ [9].

An interesting fact about the algebra (2.8a) is that, the Virasoro algebra
 with a
central extension
can be obtained from it as a infinite linear combination of the form
$L_N=\sum c_N^{\underline n}
L_{\underline n}$. The exact form of the constant coefficients
$c_N^{\underline n}$ can be found in
Ref.~[6].

A super extension of ${\rm SDiff}(T^2)$ does exist [9,4]. For completeness,
we
reproduce it here:
$$
     \eqalign{
[L_{\underline n}, L_{\underline m}] &= {\underline n}\times {\underline m}
 L_{{\underline n}+{\underline m}},  \cr
[L_{\underline n}, G_{\underline m}] &={\underline n}\times {\underline m}
G_{{\underline n}+{\underline m}},  \cr
\{G_{\underline n}, G_{\underline m}\} &= L_{{\underline n}+{\underline m}},
\cr}\eqno(2.9)
$$
where $G_{\underline n}$ are, of course, the superpartners of the bosonic
generators
$L_{\underline n}$. Note that, not merely is there no central extension
possible in the
$\{ G,G\}$
anticommutator, but also the central term that could be present in the
$[L,L]$
commutator in the
bosonic case must now be absent by Jacobi identities.

 Another simple example of a compact, orientable surface is a 2-sphere
$S^2$. As basis
functions it is
natural to choose the spherical harmonics $Y^\ell_m(\theta,\phi)$. They
furnish a
representation of
the rotation group SO(3), and they form a complete and orthonormal set.
Let us
furthermore take the
symplectic structure to be $\Omega^{ab}={\rm sin}^{-1}\theta\epsilon^{ab}$,
and expand
the parameters
and the generators of ${\rm SDiff}(S^2)$ as $ \Lambda =\sum_{\ell,m}
\Lambda^\ell_m
Y^\ell_m$ and $L_\Lambda =\sum_{\ell,m} \Lambda^\ell_m L^\ell_m$.
Since there exist no harmonic 1-forms
on $S^2$, there will be no non-trivial central extension
and the generators $L^\ell_m $ obey the following classical
${\rm SDiff}(S^2)$ algebra
$$
    [L^\ell_m, L^j_n]= c^{\ell j}_k(m,n) L^k_{m+n}  \eqno(2.10)
$$
where  $c^{\ell j}_k(m,n) $ are the structure constants, which are
essentially the
Clebsch-Gordan
coefficients of $SO(3)$ and can be written as
$$
c^{\ell j}_k(m,n) =\int d\theta d\phi \bigg(\epsilon^{ab}
\partial_b Y^\ell_m
\partial_a Y^j_n\bigg) Y^k_{-m-n}  \eqno(2.11)
$$

If we take the surface $\Sigma$ to be a 2-sphere with north and south
poles
removed, $S^2\backslash \pm 1$, which is topologically equivalent to a
cylinder, then
there will be a
nontrivial central extension. Its form is somewhat complicated, and it can
 be found in
Ref.~[19].

We next consider some examples of noncompact surfaces. In the  case of
2-hyperboloid, $H^2$, we can expand the generators of ${\rm SDiff}(H^2)$ in
a basis
closely related to the
spherical harmonics. In such a basis the structure constants of
${\rm SDiff}(H^2)$ are
essentially the
Clebsh-Gordan coefficients of $SO(2,1)$.  Since $SO(2,1)$ is contained as a
subalgebra of
${\rm SDiff}(H^2)$, if we identify it with the Lorentz group in $2+1$
dimensions, then the possibility
of interpreting ${\rm SDiff}(H^2)$ as a higher spin algebra in $2+1$
dimensions arises. If we take two
copies of ${\rm SDiff}(H^2)$, we can then identify the subgroup
$SO(2,1)\oplus SO(2,1)$
with the anti
de Sitter group in $2+1$ dimensions, and thus obtain its infinite
dimensional extension as a higher
spin algebra [20]. This algebra is very similar to the infinite dimensional
extension of
the AdS group
$SO(3,2)$ in four dimensions obtained in [21].

$H^2$ is topologically equivalent to a disk and hence its first Betti
number is zero. Consequently, ${\rm SDiff}(H^2)$ does not admit a nontrivial
central
extension. However,
if we remove the origin, the resulting surface $H^2\backslash \{ 0 \}$ is
topologically
equivalent to a cylinder and the associated area-preserving diffeomorphism
algebra will
have a
nontrivial central extension. In using the general formula (2.5) to compute
 the central
extension,
however, care must be exercised in choosing an integration measure such that
the relevant
integrals
are well defined. In what follows we shall mainly concentrate on the
classical
area-preserving
diffeomorphisms.

The surfaces of interest are those whose area preserving diffeomorphisms
can be
represented in
a basis  such that the algebra of area-preserving diffeomorphisms are
manifestly an
infinite dimensional extension of the Virasoro algebra, i.e. such that the
generators of
${\rm SDiff}(\Sigma)$ decompose in a simple manner under the Virasoro
group. To this end
 let us first
consider the area-preserving diffeomorphisms of the upper half plane
$R^2_+$, with
coordinates $x,y$
[2]. A choice of basis which would be suitable for expansions of square
integrable
functions on the
plane, would be the unitary representation of the Euclidean group in two
dimensions,
$E_2$. These are
essentially the Bessel functions. However, in such a basis the embedding of
the Virasoro
algebra
would not be manifest. To make the connection with the Virasoro algebra
manifest, a more
suitable
choice of basis is $y^{\ell+1}x^{\ell+m+1}$ where $\ell, m$ are integers
[2,22]. To avoid
the singularities at the origin, we may remove the origin, and thus
consider
$R^2\backslash \{ 0\}$,
which is topologically equivalent to a cylinder. Note that, the basis
functions diverge
as
$x,y\rightarrow \infty$, and they are not orthonormal either. However,
as we will be
interested in
functions which admit a Taylor expansion we shall not be concerned about
these properties
 of the
basis functions. Let us proceed by choosing the symplectic structure to
be
$\Omega^{ab}=\epsilon^{ab}$, and expanding the generators and the parameters
 as $\Lambda
=-\sum_{\ell,m} \Lambda^\ell_m y^{\ell+1}x^{\ell+m+1}$ and
$L_\Lambda =\sum_{\ell,m} \Lambda^\ell_m
v^\ell_m$. In this basis, the generators $v^\ell_m$ take the form
$$
  v^\ell_m= y^\ell x^{\ell+m}\big[ -(\ell+1) x{\partial\over \partial x}+
(\ell+m+1)
y{\partial\over \partial y}\big], \quad\quad m\ge -\ell-1   \eqno(2.12)
$$
and they obey the following ${\rm SDiff}(R^2)$ algebra
$$
[v^\ell_m, v^j_n] = [(j+1)(\ell+m+1)-(\ell+1)(j+n+1)] v^{\ell+j}_{m+n}
\eqno(2.13)
$$
Clearly  $v^0_m$ generate the Virasoro algebra without central extension:
$$
   [v^0_m, v^0_n]=(m-n)v^0_{m+n}  \eqno(2.14)
$$
Furthermore, the generators $v^\ell_m$  form a representation of the
Virasoro algebra
since
$$
[v^0_n, v^\ell_m] = [(\ell+1)n-m] v^\ell_{m+n} \eqno(2.15)
$$
This is to be compared with the commutation rule between the Virasoro
generators $L_n$
and the
Fourier modes of a spin $s$ conformal field $w^s_m$ given by
$$
    [L_n, w^s_m]=[(s-1)n-m]w^s_{m+n}  \eqno(2.16)
$$
Therefore the generators $v^\ell_m$ can be viewed as the Fourier
modes (labelled by $m$) of a conformal field of conformal spin $(\ell+2)$.

 Another choice of basis functions for ${\rm SDiff}(R^2)$ considered in
Ref.~[23] is
$x^{s+m}y^{s-m}$. Using the expansions $\Lambda
=-\sum\Lambda^s_m x^{s+m}y^{s-m}$ and $L_\Lambda=\sum \Lambda^s_m v^s_m$,
one finds that
${\rm SDiff}(R^2)$ algebra in this basis takes the form
$$
 [v^s_m, v^t_n]=\big[(t-n)(s+m)-(s-m)(t+n)\big] v^{s+t-1}_{m+n}  \eqno(2.17)
$$
In this basis $\ft12 v^1_n$ obeys the Virasoro algebra, and by commuting it
with $v^s_m$
we find
that $v^s_m$ can be viewed as the Fourier modes (labelled by $m$) of a
conformal field of
spin
$(s+1)$. Note that in order to avoid negative powers of $x$ and $y$ in the
basis
functions, we
must keep the generators $v^s_m$ with  $-s\ge m \le s$. Note also that in
order to
have integer
powers of $x$ and $y$, $s,\ n$ must be both integers or both half integers.

There are other choices of basis functions which give rise to algebras with
a somewhat
different interpretation of the generators as Fourier modes of conformal
ields.
For example, using
the polar coordinates $r,\theta$, consider the basis functions $r^{\ell+2}
e^{im\theta}$.
Furthermore, choosing the symplectic structure to be $\Omega^{ab}=r^{-1}
\epsilon^{ab}$,
and using the
expansions $\Lambda=-i\sum\Lambda^\ell_m r^{\ell+2}e^{im\theta}$ and
$L_\Lambda =\sum_{\ell,m} \Lambda^\ell_m v^\ell_m$, where $\ell,m$ are
integers, we find that ${\rm SDiff}(R^2\backslash\{0\})$ in this basis takes
the form
$$
[v^\ell_m, v^j_n]=[(j+2)m-(\ell+2)n]v^{\ell+j}_{m+n},  \eqno(2.18)
$$
where $\ell\ge -1$ and $-\infty<m<\infty$. In this basis, ${1\over 2}v^0_m$
obey the
Virasoro algebra.
Using the notation $L_m={1\over 2}v^0_m$, from (2.18) we obtain
$$
  [L_m, v^\ell_n]=\Bigg[{1\over 2}(\ell+2) -n\Bigg] v^\ell_n   \eqno(2.19)
$$
Compared with (2.16), this implies that $v^\ell_n$ can be viewed as the
Fourier modes
(labeled by
$n$) of a conformal field of spin $((\ell+4)/2$. We shall call this algebra
the
{\it twisted
$w_\infty$ algebra}.

Finally, let us describe a somewhat more convenient choice of basis to
describe the
algebra of area-preserving diffeomorphisms of a surface with the topology
of a cylinder,
$R\times S^1$. Let the coordinates of the cylinder be $0\le x\le 2\pi$ and
$\infty < y
< \infty$.  A suitable choice of basis functions is
$y^{\ell+1}e^{imx}$ [8]. Choosing the symplectic structure to be
$\Omega^{ab}=\epsilon^{ab}$ and
using the expansions $\Lambda=-i\sum \Lambda^\ell_m y^{\ell+1}e^{imx}$ and
$L_\Lambda =\sum_{\ell,m} \Lambda^\ell_m v^\ell_m$, we find that
${\rm SDiff}(R\times S^1)$ algebra takes the form
$$
[v^i_m, v^j_n] = [(j+1)m-(i+1)n] v^{i+j}_{m+n}, \eqno(2.20)
$$
where $\ell\ge -1$ and $-\infty<m<\infty$. Evidently $v^0_m$ obeys the
Virasoro algebra
and from the
commutator
 $$
[v^0_n, v^\ell_m] = [(\ell+1) n-m] v^\ell_{m+n} \eqno(2.21)
$$
we see that $v^\ell_m$ are the Fourier modes (labelled by $m$) of a
conformal field of
spin
$(\ell+2)$. In particular $v^{-1}_m$ has spin 1. This algebra is usually
referred to as
$w_{1+\infty}$ algebra. If we exclude the spin 1 generator, we still have
a closed
algebra, known as
the $w_\infty$ algebra.

Experience with Virasoro algebra suggests that in order to have a nontrivial
unitary
representation
of $w_{1+\infty}$ it should admit a central extension. Starting directly
from (2.15)
and searching
for central extensions by means of checking the Jacobi identities, one
finds that it
is allowed only
in the spin 2 sector. Denoting the central extension that arises in the
commutator of
$v^i_m$ with
$v^j_n$  by $c^{ij}(m,n)$, it takes the familiar form
$$
c^{ij}(m,n)={c\over 12}(m^3-m)\delta^{i,0}\delta^{j,0}\delta_{m+n,0},
\eqno(2.22)
$$
where $c$ is an arbitrary constant. An extension of  $w_{1+\infty}$ which
does
contain a central extension in all spin sectors exists, and it is referred
to as
$W_{1+\infty}$.
It turns out that $w_{1+\infty}$ algebra can be viewed as a contraction of
the latter.
Alternatively,
$w_{1+\infty}$ can be interpreted as the classical limit of the quantum
algebra
$W_{1+\infty}$.

$w_{1+\infty}$ algebra admits two natural subalgebras. One of them, which
we shall
denote by
$w_{1+\infty}^+$ has the generators $v^\ell_m$ with the restriction
$\ell\ge -\ell-1,\ \infty\le m\le \infty $, and the other one,
$w_{1+\infty}^-$ generated by
$v^\ell_m$ with the restriction  $\ell\le -\ell-1,\ \infty\le m\le \infty$.
Another
useful
subalgebras is the Cartan subalgebra. There are a number of ways of
choosing it. For
example,
$v^\ell_0$ are infinitely many mutually commuting generators, thus forming
the Cartan
subalgebra
\footnote{$^\dagger$}{It is interesting
to note that, since in particular $[v^0_0,v^n_0]=0$, and $v^0_0$ is the
usual Hamiltonian
$H=L_0=v^0_0$, we can interpret $v^n_0,\ n\ge 1 $ as infinitely many
conserved quantities
that
commute with the Hamiltonian.}.
Another set of mutually commuting generators are $v^\ell_{-\ell-1}$ and
$v^\ell_{\ell+1}$. These
algebras will play a role when we discuss the nonlinear realizations of
$w_{1+\infty}$.
\bigskip
\noindent{\bf 3. Generalizations of the $w_\infty$ Algebras}
\bigskip
There are a number of extensions of the area-preserving diffeomorphisms.
Among them are
the $N=1$
[24] and $N=2$ [22] supersymmetric extensions of $w_\infty$. The $N=2$
super $w_\infty$
algebra takes
the form [22,25]
$$
\eqalign{
   [v^i_m, v^j_n] &=\big[ (j+1)m-(i+1)n\big]
v^{i+j}+{c\over 8}(m^3-m) \delta^{i,0}\delta^{j,0}\delta_{m+n,0} \cr
   [v^i_m, J^{j-1}_n] &= \big[jm-(i+1)n\big] J^{i+j-1}_{m+n} \cr
\{{\bar G}^\alpha_r, G^\beta_s \} &=2 v^{\alpha +\beta}_{r+s}
-2\big[(\beta+\half )r-(\alpha+\half)s\big]J^{\alpha+\beta-1}_{r+s}+
{c\over
2}(r^2-{1\over 4})\delta^{\alpha,0}\delta^{\beta,0}\delta_{r+s,0} \cr
[v^i_m, G^\alpha_r] &= \big[(\alpha+\half )m-(i+1)r\big]G^{\alpha+i}_{m+r}
\cr
[v^i_m, {\bar G}^\alpha_r] &=
\big[(\alpha+\half )m-(i+1)r\big]{\bar G}^{\alpha+i}_{m+r} \cr
[J^{i-1}_m, G^\alpha_r] &=G^{i+\alpha}_{m+r} \cr
     [J^{i-1}_m, {\bar G}^{\alpha}_r] &=-{\bar G}^{\alpha+i}_{m+r}.\cr
    [J^{i-1}_m, J^{j-1}_n] &= {c\over 2} m
                     \delta^{i,0}\delta^{j,0}\delta_{m+n,0}, \cr }\eqno(3.1)
$$
where the notation is self explanatory. In fact, this is the algebra of
symplectic
diffeomorphisms on
a $(2,2)$ superplane, i.e. a plane of two bosonic and two fermionic
dimensions
[24]. The $N=1$ super $w_\infty$ algebra can be obtained from the above
algebra by truncation, or directly as an algebra of the symplectic
diffeomorphisms of a $(2,1)$ superplane [24,22]. For $i=j=\alpha=\beta=0$,
 the
algebra (3.1) reduces to the well known $N=2$ superconformal algebra.

Yet another extension is called the topological
$w_\infty$ algebra denoted by $w^{\rm top}_\infty$ [26]. It is obtained
 from the
$N=2$ super
$w_\infty$ algebra by a twisting procedure introduced by Witten [27].
The idea is to
identify one of
the fermionic  generators as the
nilpotent BRST charge Q, and to define  bosonic generators which can be
written
in the form ${\hat v}^i_m =\{ Q,{\rm something} \}$. This is the higher-spin
generalization of the property that holds for the energy-momentum tensor of
a
topological field theory. A suitable candidate for the BRST charge is
$ Q=-{\bar G}^0_{-{1\over 2}}$. We then define the generators of
$w_{\infty}^{\rm top}$ to be $G^i_{m+{1\over 2}}$ and define ${\hat v}^i_m$
as
${\hat v}^i_m=-\{Q,G^i_{m+{1\over2}}\}$. It can be easily shown that these
 generators
obey the
algebra
$$
\eqalign{
[{\hat v}^i_m,{\hat v}^j_n]&=\Big[(j+1)m-(i+1)n\Big]
{\hat v}^{i+j-\ell}_{m+n}\cr
[{\hat v}^i_m,G^j_{n+\ft12}]&=\Big[(j+1)m-(i+1)n\Big]
G^{i+j-\ell}_{m+n+\ft12}\cr
\{G^i_{m+\ft12},G^j_{n+\ft12}\}&=0\cr}\eqno(3.2)
$$
Note that the structure constants for $[{\hat v}^i_m,G^j_{n+\ft12}]$ are
the
same as those for $[{\hat v}^i_m,{\hat v}^j_n]$, and that the algebra is
centerless.
A field
theoretic realization of $W^{\rm top}_{\infty}$ is given in Ref.~[26].

Another interesting extension of $w_\infty$ involves a Kac-Moody sector.
It can be obtained directly by a suitable contraction of the $W_{1+\infty}$
algebra with
$SUN)$
symmetry found in Ref.~[28]. The generators of the algebra are $v^i_m$ and
$J^{i,a}_m$
where $i$ is an
integer such that $i\ge -1$ and $a=1,...,N^2-1$ labels the adjoint
 representation of
$SU(N)$ and they
have the commutation relations
$$
\eqalign{
      [v^i_m, v^j_n] &= [(j+1)m-(i+1)n] v^{i+j}_{m+n}+
          {c\over 12}(m^3-m)\delta^{i,0}\delta^{j,0}\delta_{m+n,0}  \cr
    [v^i_m, J^{j,a}_n] &= [(j+1)m-(i+1)n] J^{i+j,a}_{m+n} \cr
[J^{i,a}_m, J^{j,b}_n] &= {1\over 2}f^{abc} J^{i+j+1,c}_{m+n}+
                   {1\over 16} km\delta^{i+1,0}\delta^{j+1,0}\delta_{m+n,0}
\cr}\eqno(3.3)
$$
where $f^{abc}$ are the structure constants of $SU(N)$, and the central
extension, $c$, of the
Virasoro algebra, and the level, $k$, of the Kac-Moody algebra are related
to each other by
the Jacobi identity: $c=Nk$. The generators of the above algebra (without
 central extension)  can be
represented as follows
$$
    \eqalign{
    v^\ell_m  &=-iy^\ell e^{imx}\Big[(\ell+1){\partial\over \partial x}
                                             -imy{\partial\over
\partial y}\Big]  \cr
J^{\ell,a}_m  &= -it^ay^{\ell+1}e^{imx},  \cr}\eqno(3.4)
$$
where $t^a$ are the generators of $SU(N)$. It would be interesting to find a
field theoretic
realization of this algebra.

There also exists an infinite dimensional generalization of $w_\infty$
algebra which is
related
to the symplectic diffeomorphisms in four dimensions. The generators are
labelled as
$v_m^{\ell,\vec k}$ where ${\vec k}=(k_1,k_2)$ and they obey the algebra
[29]
$$
[V_m^{\ell,\vec k},
V_n^{j,\vec \ell}]=[(j+1)m-(\ell+1)n]
V^{\ell+j,{\vec k+\vec \ell}}_{m+n}+
{\vec k}\times {\vec \ell}\ V^{\ell+j+1,{\vec k+\vec \ell}}_{m+n},
\eqno(3.5)
$$
Another infinite dimensional extension of the $w_\infty$ algebra is the
loop algebra of
$w_\infty$.
For example, the loop algebra of ${\rm SDiff}(R^2)$ in the basis (2.17) is
$$
[v^s_m(\sigma), v^t_n(\sigma')]=
 \big[(t-n)(s+m)-(s-m)(t+n)\big] v^{s+t-1}_{m+n}(\sigma)
         {\p\over \p\sigma}\delta (\sigma-\sigma')  \eqno(3.6)
$$

This concludes the brief survey of some of the salient features of
the area-preserving diffeomorphisms, $w_{1+\infty}$ algebras and their
generalizations.
We now turn to their field theoretic realizations.
\bigskip
\noindent{\bf 4. $w_\infty$ Gravity and Supergravity}
\bigskip
Before we describe field theoretic realizations of $w_\infty$ symmetry, it
is useful to
review
a simple field theoretic realization of the Virasoro symmetry, and to
formulate it in a
language that
lends itself readily to a $w_\infty$ generalization. To this end consider
the Lagrangian
$$
{\cal L} =-{1\over 4}{\sqrt -h}h^{ij}\partial_i\phi
\partial_j\phi,   \eqno(4.1)
$$
where $h^{ij}$ is the inverse of the worldsheet metric
$h_{ij},\ (i,j=0,1),\ h=det h_{ij}$ and $\phi$ is a real scalar. This
Lagrangian clearly possesses the 2D diffeomorphism and Weyl symmetries.
It is convenient to parametrize the metric as follows [25,14]
$$
    h_{ij}=\Omega \left(\matrix{2h_{++}&1+h_{++}h_{--}\cr
                         1+h_{++}h_{--}&2h_{--}\cr}\right),  \eqno(4.2)
$$
where $\Omega$ is an arbitrary function which drops out in the action, and
the light-cone
coordinates
are defined by $x^{\pm}={1\over \sqrt 2}(x^0\pm x^1)$. In terms of these
 variables the
Lagrangian
(4.1) becomes
 $$
{\cal L}={1\over 2}(1-h_{++}h_{--})^{-1}
(\partial_+\phi-h_{++}\partial_-\phi)
(\partial_-\phi-h_{--}\partial_+\phi). \eqno(4.3)
$$
It turns out to be very useful to rewrite this Lagrangian in the following
first order form
$$
{\cal L}=-{1\over 2}\partial_+\phi \partial_-\phi -J_+J_- +J_+\partial_-\phi
+J_-\partial_+\phi-{1\over 2}h_{--}J_+^2-{1\over 2}h_{++}J_-^2, \eqno(4.4)
$$
where $J_{\pm}$ are auxiliary fields which obey the field equations
$$
     \eqalign{
              J_+ &=\partial_+\phi-h_{++}J_-, \cr
              J_- &=\partial_-\phi-h_{--}J_+, \cr}\eqno(4.5)
$$
These equations define a set of nested covariant derivatives [30]. Solving
for
$J_{\pm}$ and substituting into (4.4) indeed yields (4.3). Thus the two
Lagrangians are classically equivalent, though in principal they may be
quantum
inequivalent. The action of the Lagrangian (4.4) is invariant under 2D
diffeomorphism transformations, with a general parameter $k_+(x^+,x^-)$,
given
by [11]
$$
\eqalign{
    \delta \phi &=k_+J_-  \cr
    \delta h_{++}&=\partial_+k_+-h_{++}\partial_-k_++k_+\partial_-h_{++} \cr
     \delta h_{--} &=0  \cr
     \delta J_- &=\partial_-(k_+J_-)   \cr
       \delta J_+ &=0,  \cr}\eqno(4.6)
$$
and 2D diffeomorphisms with parameters $k_-(x^+,x^-)$, which can be
obtained from the above transformations by changing  $+\leftrightarrow -$
everywhere.

The $W$ symmetric generalization of (4.4) is now remarkably simple. With
the
further generalization to the case in which the fields $\phi$ and $J_{\pm}$
take their values in the Lie algebra of $SU(N)$ the answer can be written as
follows [11]
$$
  \eqalign{
           {\cal L}=& {\rm tr}\ \big(-{1\over 2}\partial_+\phi
\partial_-\phi-J_+J_- + J_+\partial_-\phi + J_-\partial_+\phi\big) \cr
   &-\sum_{\ell\ge 0}{1\over {\ell+2}}A_{+\ell}{\rm tr}\ J_-^{\ell+2}
 -\sum_{\ell\ge 0}{1\over {\ell+2}}A_{-\ell}{\rm tr}\ J_+^{\ell+2}.\cr}
\eqno(4.7)
$$
 Note that $A_{0+}=h_{++}$ and $A_{0-}=h_{--}$. The equations of motion
for the
auxiliary fields now reads
$$
     \eqalign{
              J_+ &=\partial_+\phi-\sum_{\ell\ge 0}A_{+\ell}J_-^{\ell+1},
 \cr
       J_- &=\partial_-\phi-\sum_{\ell\ge 0}A_{-\ell}J_+^{\ell+1}. \cr}
\eqno(4.8)
$$
The Lagrangian (4.7) possesses the $w_\infty$ symmetry with parameters
$k_{+\ell}(x^+,x^-)$ that generalize (4.6) as follows [11]
$$
\eqalign{
\delta\phi&=\sum_{\ell\ge -1}k_{+\ell}J_-^{\ell+1}\cr
\delta A_{+\ell}&=\partial_+ k_{+\ell} -
\sum_{j=0}^\ell [(j+1)A_{+ j}\partial_- k_{+(\ell-j)} -
(\ell-j+1)k_{+(\ell-j)} \partial_- A_{+ j}]\cr
\delta A_{-\ell}&=0  \cr
\delta J_-&=\sum_{\ell \ge -1}\partial_-\big[k_{+\ell}(J_-)^{\ell+1}\big]\cr
  \delta J_+&=0,  \cr}\eqno(4.9)
$$
and $W$ transformations with parameters $k_{+\ell}(x^+,x^-)$ which can be
obtained from above by the replacement $+\leftrightarrow -$ everywhere.
It is important to note that we must set
$$
    k_{-1\pm}=-{1\over N}\sum_{\ell\ge 1} k_{\pm\ell}{\rm tr}\
J_{\mp}^{\ell+1},
\eqno(4.10)
$$
to ensure the tracelessness $\delta \phi$ and $\delta J_{\pm}$ in the
transformation rules above. The Lagrangian (4.8) has also Stueckelberg type
shift
symmetries which
arise due to the fact that for $SU(N)$, only $(N-1)$ Casimirs of the form
${\rm tr}(J_{\pm})^{\ell+2}$ are really independent, while the rest can be
factorize into products of these Casimirs. For a further discussion of
these symmetries,
see
Ref.~[11].

There exists an interesting chiral truncation of the $w$ gravity theory
discussed above. It is achieved by setting $A_{+\ell}=0$. In that case from
(4.10) we have $J_+=\partial_+\phi$ and $J_-=\partial_-\phi
-\sum_{\ell\ge 0}A_{-\ell}{\rm tr}(\partial_+\phi)^{\ell+1}$. In this case,
it
is more convenient to work in second order formalism. Thus, substituting for
$J_{\pm}$ into the Lagrangian (4.7), we obtain [11]
$$
  {\cal L}= {1\over 2}{\rm tr}\partial_+\phi
\partial_-\phi -\sum_{\ell\ge 0}{1\over {\ell+2}}
A_{\ell}{\rm tr}(\partial_+\phi)^{\ell+2}, \eqno(4.11)
$$
where we have used the notation $A_{-\ell}=A_\ell$. This  Lagrangian has the
following  symmetry [11]
$$
\eqalignno{
\delta\phi&=\sum_{\ell\ge -1}k_\ell(\partial_+\phi)^{\ell+1}  &(4.12)\cr
\delta A_\ell&=\partial_- k_\ell -
\sum_{j=0}^{\ell+1} [(j+1)A_j\partial_+ k_{\ell-j}-
(\ell-j+1)k_{\ell-j} \partial_+ A_j]  &(4.13)\cr}
$$
The Lagrangian (4.11) has also the appropriate Stueckelberg symmetry. Using
this symmetry one can obtain [11] the chiral $W_3$ gravity of Ref.~[31].
Note
that the interaction term in this Lagrangian has the form of a  gauge field
$\times$  conserved current ${1\over {\ell+2}}{\rm
tr}(\partial_+\phi)^{\ell+2}.$ It is important to note that the OPE of these
currents do not form a closed algebra, while they do close with respect to
Poisson bracket. Hence, the $w_\infty$ symmetry described above is a
classical symmetry, as expected.

The $w_\infty$ gravity with $N=2$ super $w_\infty$ symmetry has also been
constructed [32]. For
readers convenience we summarize the result of Ref.~[32] for chiral $N=2$
super
$w_\infty$ here.
Consider two real scalar superfields  $\phi$ and ${\bar \phi}$. Let the
superspace
coordinates be
$Z=(z,\theta)$, and define the covariant derivatives
$D=\partial_\theta-\theta\partial$
and ${\bar
D}=\partial_{\bar \theta}-{\bar \theta}{\bar \partial}$. The currents which
classically
generate
the $N=2$ super $w_\infty$ algebra are [32]
$$
  \eqalign{
w^\ell &=D\phi(\partial\phi)^{\ell+1}D{\bar \phi},  \quad\quad {\rm for }
\quad\quad
\ell=-1,0,1,2,...\cr &=(\partial\phi)^{\ell+{3\over 2}}D{\bar \phi}+\half
D\Big[D\phi(\partial\phi)^{\ell+{3\over 2}}D{\bar \phi}\Big] \quad\quad
{\rm for }
\quad\quad
\ell=\half, {3\over 2},...  \cr}\eqno(4.14)
$$
In terms of these currents the chiral $N+2\ w_\infty$ supergravity action
takes the
form [32]
$$
S=\int d^2 Z\Big[D\phi{\bar D}{\bar \phi}+\sum_{\ell=-1}^\infty A_\ell
w^\ell\Big]
\eqno(4.15)
$$
The transformation rules for the matter fields are [32]
$$
\eqalign{
  \delta \phi &=\sum_{\ell=-1,0,...}^\infty k_\ell D\phi
(\partial\phi)^{\ell+1}+
\sum_{\ell=-\half,\half ,...}\Big[ k_\ell(\partial\phi)^{\ell+{3\over 2}}
-\half Dk_\ell D\phi
(\partial\phi)^{\ell+\half }\Big] \cr
\delta {\bar \phi} &= \sum_{\ell=-1,0,...}^\infty
\Big\{ -k_\ell(\partial\phi)^{\ell+1}
D{\bar \phi}
- (\ell+1)D\big[k_\ell D\phi(\partial\phi)^\ell D{\bar \phi}\big] \Big\}
\cr
   &+\sum_{\ell=-\half ,\half ,...}^\infty \Big\{ \half Dk_\ell
(\partial\phi)^{\ell+\half }-
    (\ell+{3\over 2}) D\big[ k_\ell (\partial\phi)^{\ell+\half}
D{\bar \phi}\big] \cr
   &\quad\quad\quad\quad\quad\quad +\half (\ell+\half )D\big[ Dk_\ell D\phi
              (\partial\phi)^{\ell-\half} D{\bar \phi}\big]\Big\}\cr}
\eqno(4.16)
$$
The gauge fields transform as follows [32]
$$
\eqalign{
\delta A_\ell &= {\bar D}k_\ell+\sum_{j=\ft12 ,1,...}
\Big[(\ft12 -j)A_j\p k_{\ell-j+\ft52}\cr
  &+\ft12 (-1)^{2j}DA_jDk_{\ell-j+\ft52}+(\ell-j+3)\p A_j k_{\ell-j+\ft52}
\Big] \cr}
\eqno(4.17a)
$$
for $ \ell=-\half,\half,...$, and
$$
\eqalign{
\delta A_\ell &= {\bar D}k_\ell-2\sum_{j=1,2,...}^{\ell+\ft32}A_jk_{\ell-j+
\ft52}+
\sum_{j=\ft12 ,\ft32,...}\Big[(\ft12 -j)A_j\p k_{\ell-j+\ft52}\cr
&+\ft12(-1)^{2j}DA_jDk_{\ell-j+\ft52}+(\ell-j+3)\p A_j k_{\ell-j+\ft52}\Big]
\cr}
\eqno(4.17b)
$$
for $\ell=-1,0,...$  The spin $\ft12$ transformations can be included as in
Ref.~[32].
The
nonchiral version of this theory has also been constructed [32].

Going back to the bosonic version of nonchiral $w_\infty$ gravity, an
interesting
geometric
formulation of it has been given by Hull [14]. Consider the case of a
single real scalar
$\phi$. According to Ref.~[14], the symmetry algebra of nonchiral
$w_\infty$ gravity
is a subalgebra
of the more general set of transformations
$$
\delta\phi=\sum_{n=2}^\infty \lambda^{{i_1}{i_2}\cdots
{i_{n-1}}}(x)\p_{i_1}\phi\p_{i_2}\phi\cdots\p_{i_{n-1}}\phi  \equiv
\Lambda(x^i, y_i),
\eqno(4.18)
$$
where $y_i=\p_i\phi$ and $\lambda^{i_1i_2\cdots i_{n-1}}(x) (n=2,...)$
are the
infinitesimal parameters which are symmetric tensors on the worldsheet. It
is
easy to show that these transformations satisfy the Poisson bracket algebra
on
the phase space with coordinates $x^i, y_i$. The action proposed in
Ref.~[14] is
$$
   S=\int d^2 x {\tilde F}(x,y)  \eqno(4.19)
$$
where ${\tilde F}(x,y) $ has the expansion
$$
  {\tilde F}(x,y) =\sum_{n=2}^\infty {1\over n}{\tilde h}^{i_1 i_2
\cdots i_n}(x) y_{i_1}y_{i_2}
\cdots y_{i_n}, \eqno(4.20)
$$
where ${\tilde h}^{i_1 i_2 \cdots i_n}(x)$ are gauge fields which are
tensor densities
on the
world-sheet. Invariance of this action under the transformations (4.18)
requires the
imposition of
the following constraint on the parameter $\Lambda(x,y)$ [14]
$$
\epsilon_{ik}\epsilon_{jl}{\p^2\Lambda\over \p y_i\p y_j}
{\p^2{\tilde F}\over \p y_k\p y_l}=0  \eqno(4.21)
$$
The expansion of this equation gives an infinite number of algebraic
constraints on the
parameters. The first such constraint is $\lambda^{ij}{\tilde h}_{ij}=0$.
Other
constraints relate
the trace of a rank $\ge 3$ parameter to the products of lower rank
 parameters and
gauge fields. The
invariance of the action under the transformations (4.18), in addition to
the
constraint (4.21),
also requires that the gauge fields transform as [14]
$$
 \eqalign{
   \delta {\tilde h}^{i_1\cdots i_p}=& \sum_{n=2}^p
\Bigg[-{(p-n+1)(n-1)\over (p-1)}\p_j\big(
{\tilde h}^{j(i_1\cdots i_{n-1}}\lambda^{i_n\cdots i_p)}-
\lambda^{j(i_1\cdots i_{p-n}}
{\tilde h}^{i_{p-n+1}\cdots i_p)}\big)\cr
&-(n-1){\tilde h}^{j(i_1\cdots i_{n-1}}\p_j\lambda^{i_n\cdots i_p)}+(p-n+1)
\lambda^{i_1\cdots
i_{p-n+1}} \p_j{\tilde h}^{i_{p-n+2}\cdots i_p)j}\Bigg]  \cr}  \eqno(4.22)
$$
It turns out that a gauge invariant condition can be imposed on the gauge
fields which
reduces
at the linearized level to the constraint present in $w_\infty$ gravity.
This constraint
 is [14]
$$
{\rm det}\ \Bigg({\p^2{\tilde F}(x,y)\over \p y_i \p y_j}\Bigg)=-1
\eqno(4.23)
$$
 The expansion of this equation gives an infinite number of algebraic
constraints on the
gauge
fields. The first two constraints are ${\rm det}\ \big( {\tilde h}^{ij}\big)
=-1$ and
${\tilde h}^{ijk}{\tilde h}_{ij}=0$. Other constraints relate the trace of
rank$\ge 4$
gauge fields
to the products of lower rank gauge fields.  The procedure for solving the
constraints
(4.21) and
(4.22) in terms of unconstrained gauge fields and gauge parameters, as well
 as the
 attendant
generalized Weyl symmetries  have been outlined in Ref.~[14].

The equation (4.23) has a nice geometrical interpretation. With an
appropriate
identification of ${\tilde F}(x,y)$ with a Kahler potential, (4.23) can be
interpreted
as the
Monge-Ampere equation for a Kahler metric of a self-dual four-manifold [14].
 Other
relations between
$w_\infty$ and self-dual geometry have  been discussed in [7,16]
\bigskip
\noindent{\bf 5. Nonlinear Realizations of $w_\infty$}
\bigskip
The field theoretic realization of $w_{1+\infty}$ in terms of a single real
scalar
as given in
(4.12) can be understood within the framework of nonlinear realizations [11].
Denoting the
coordinates
of a cylinder by $(x^+, y)$, the scalar field $\phi(x^+,x^-)$ can be
considered as
parametrizing the coset space $w_{1+\infty}/w_\infty $. The generators of
this coset are
$v^{-1}_m=-ime^{im\theta}\p_y$, and therefore we can choose the coset
representative
to be
$e^{\p_+\phi\p_y}$. If we denote the coset generators by $K$, and the
subalgebra generators by $H$, then one can see from (2.19) that the
structure of the
algebra is
$[H,H]\subset H$, $[H,K]\subset H+K$, $[K,K]=0$.  Thus the coset
$w_{1+\infty}/w_\infty$ is not a symmetric space, and the $K$ generators do
not even form a linear representation of $H$ (except when the subalgebra
element lies in the Virasoro subalgebra).  Although we can apply the
general theory of non-linear realizations to this situation, we should not
be surprised to find that some of the features are non-standard. In
particular, the action of $H$ on the coset will in general be non-linear.

     Acting on the coset representative with a general $G$ transformation
$g$, one has
$$
g e^{\p_+\phi\partial_y}=e^{\p_+\phi\partial_y} h,\eqno(5.1)
$$
where $h$ is an element of the subalgebra $H$. For an infinitesimal
transformation $g=1+\delta g$, one therefore finds
$$
\p_+ \delta\phi =\Big(e^{-\p_+\phi\partial_y}\ \delta g\
e^{\p_+\varphi\partial_y}\Big)_{G/H}
\eqno(5.2)
$$
Taking $\delta g=k_\ell(x^+) y^{\ell+1}$ we find that
$\p_+ \delta\phi=\p_+\big[k_\ell(\p_+\phi)^{\ell+1}\big]$. Hence we have
$$
\delta \phi=k_\ell(x^+)(\p_+\phi)^{\ell+1}  \eqno(5.3)
$$
which is indeed the global $w_{1+\infty}$ transformation discussed in the
 previous
section. This
result can also be derived in the Poisson bracket language. Corresponding
to (5.2)
we have
$$
\eqalign{
   \delta\phi=&\big(e^{-Ad_\phi}\ \delta g\big)\mid_{y=0} \cr
  &=\big( \delta g+\{\phi,\delta g\}+\ft1{2!}\{\phi,\{\phi,\delta g\}\}+
                                       \cdots\big)\mid_{y=0}   \cr}
\eqno(5.4)
$$

 Setting $y=0$ amounts to restriction to the coset direction. To generalize
this
construction to the
case of the non-chiral bosonic $w_\infty$ model in the light cone gauge, we
consider two copies of the
cylinder with coordinates $(x^+, y)$ and $x^-, {\tilde y})$, so that we have
 a total
of four
coordinates $(x^+, x^-, y, {\tilde y})$. The  coordinates $y$ and $\tilde y$
 play the
r\^ole of
``momenta'' conjugate to $x^+$ and $x^-$, and  so one can think of the
four-dimensional
space as
being the cotangent bundle  $T^*(\Sigma)$ of the 2-torus $\Sigma$.  The
area-preserving
transformations on the two cylinders  may be described by introducing the
symplectic form
$$
\Omega=dx^+\wedge dy-dx^-\wedge d\tilde y,\eqno(5.5)
$$
and using this to define a Poisson bracket $\{f,g\}=\Omega^{ij}\partial_i
f\,\partial_j g$ in the four-dimensional space. The symplectic
diffeomorphisms which
leave this
form invariant are  $\delta x^i=\{x^i,\delta g\}$, and the global
non-chiral $w_\infty$ transformations correspond to restricting
$\delta g$ to
have the form   $\delta g= k_{-\ell}(x^+) y^{\ell+1} + k_{+\ell}(x^-)
{\tilde y}^{\ell+1}$.  Their
action on $\phi$ is given by (5.4), where ${\rm Ad}_\phi \delta g$ is now
taken to be $\Omega^{ij}\partial_i \varphi\, \partial_j \delta g $ [11].

 It is instructive to consider some other nonlinear realizations of
$w_{1+\infty}$ involving more than a single scalar field. We have considered
elsewhere [15] the
nonlinear realization based on the coset
$w^\uparrow_{1+\infty}/{\rm Vir}^+$,
where  $w^\uparrow_{1+\infty}$ is generated by $v^\ell_m$ with
$m\ge -\ell-1$, and
${\rm Vir}^+$ is
generated by $v^0_m$ with $m\ge 0$, i.e the positive modes of the Virasoro
algebra. In
this construction we view the coordinate $x^+$ itself as a coset parameter
associated with the
Virasoro generator $L_{-1}=v^0_{-1}$. We choose the coset representative
$$
    k=e^{-x^+ v^0_{-1}}\prod_{\ell\ne 0}e^{-\phi^{\{\ell\}}}, \eqno(5.6)
$$
where we have used the notation
$\phi^{\{\ell\}} \equiv \sum_{m=-\ell-1}^\infty \phi^\ell_m v^\ell_m $.
 Note that there
are
infinitely many initial Goldstone fields $\phi^\ell_m$ corresponding to
the coset
generators
$v^\ell_m,\  \ell\ne 0$. Excluding the generator $v^{-1}_0$ which does not
correspond
to a Goldstone
field but to the coordinate $x^+$, all the remaining generators of the coset
form a
linear
representation of the subalgebra. We can construct the Cartan-Maurer form as
follows
$$
   {\cal P}=k^{-1}d k= E^0_{-1}v^0_{-1}+
\sum_{\ell\ne 0} E^\ell_m v^\ell_m + \sum_{m\ge 0}\omega^0_m v^0_m,
\eqno(5.7)
$$
where $E^\ell_m$ and $\omega^0_m$ are all 1-forms, {\it i.e.}\ $E^\ell_m=
dx^+E_m^\ell$.
 Next we look
for a maximum set of covariant constraints with which we can eliminate the
inessential
Goldstone
fields. Such a constraint is
$$
E_m^\ell = 0, \quad {\rm for}\ \ \ell\ne 0   \eqno(5.8)
$$
As a consequence of we find that
$$
\eqalignno{
E_{-1}^0 &=-dx^+,  \quad\quad\quad  \omega_m^0=0  \cr
\phi^\ell_m &=-{1\over \ell+m+1}\p_+\phi^\ell_{m-1},\quad {\rm for}\ \
m\ge -\ell  &(5.9)\cr}
$$
The last relation shows that only the Goldstone fields $\phi^\ell_{-\ell-1}$
 (the ``edge
fields'') survive the constraints, and all the other Goldstone fields can be
 expressed
in terms of their derivatives.

The transformation rules for the surviving Goldstone fields can be
derived as follows. The action of the group $w_{1+\infty}^\uparrow$ on a
coset
representative, which
we shall generically denote $e^{-\phi(x)}$, is given by
$$
        g e^{-\phi(x)} =e^{-\phi'(x')} h,  \eqno(5.10)
$$
where $h$ is an element of the divisor subgroup ${\rm Vir}^\uparrow$. For
infinitesimal transformations, we have
$$
        e^{\phi(x)+{\bar \delta}\phi(x)}(1+\delta g) e^{-\phi(x)}=
1+\delta h, \eqno(5.11)
$$
where
$$
   \eqalign{
             {\bar \delta}\phi(x) &\equiv \phi'(x')-\phi(x)  \cr
                           & =\phi'(x)+\delta x^+\partial_+\phi(x)-
\phi(x) \cr
     &\equiv \delta\phi(x)+\delta x^+\partial_+\phi(x).\cr}  \eqno(5.12)
$$
Projecting (5.11) into the coset direction yields the formula
$$
\eqalign{
    \Big(e^\phi{\bar \delta}e^{-\phi}\Big)\for_{G/H}
    &=\Big(e^\phi\delta g e^{-\phi}\Big)\for_{G/H}\cr}. \eqno(5.13)
$$
Upon the use of (5.9), the variation (5.12) simplifies to
${\bar \delta}\phi^{\{\ell\}} =\delta \phi^{\{\ell\}}-
[\phi^{\{\ell\}},v^0_{-1}]\delta x^+$.
Substituting this result in (5.13) we find that all the $\delta x^+$  terms
 cancel
pairwise except
the $-\delta x^+$ in the $v^0_{-1}$ direction. Studying (5.13) level by
level in spin,
the
transformation rules for all the Goldstone fields can now be derived. To
this end it is
useful to
consider the first two levels of dressing of $\delta g$ which we parametrize
 as follows
$$
\delta g =\sum_{\ell, m} \alpha^\ell_m v^\ell_m, \eqno(5.14)
$$
where the $\alpha^\ell_m$ are constant parameters.
Consider a spin-$\ell$ transformation with parameter $\alpha^{\{\ell\}}
\equiv
\sum_m \alpha^\ell_m v^\ell_m$. Defining
$$
e^{x^+v^0_{-1}}\alpha^{\{\ell\}}e^{-x^+v^0_{-1}}=\sum_m \beta^\ell_m (x^+)
v^\ell_m,
\eqno(5.15)
$$
we find that
$$
\beta^\ell_{m+1}(x^+)={-1\over {\ell+m+2}}\partial_+\beta^\ell_m(x^+).
 \eqno(5.16)
$$
Note that this is the same relation as that satisfied by the fields
$\phi^\ell_m$. Proceeding on to the next level of dressing, we define
$$
\sum_{\ell, m}e^{\phi^{\{-1\}}} \beta^\ell_m(x^+) v^\ell_m
e^{-\phi^{\{-1\}}}
     =\sum_{\ell, m} \gamma^\ell_m(\phi^{\{-1\}})v^\ell_m  \eqno(5.17)
$$
The field dependent parameters $\gamma^\ell_m$ can be straightforwardly
computed in
terms of
$\beta^\ell_{m+1}(x^+)$, and as a consequence one finds that
$$
\eqalignno{
\gamma^\ell_{-\ell-1} &={1\over (\ell+1)!}{\delta^{\ell+1}
\gamma^{-1}_0 \over \delta y^{\ell+1}}, &(5.18)  \cr
 \gamma^{-1}_0 &= \sum_{\ell=-1}^\infty \beta^\ell_{-\ell-1}(x^+) y^{\ell+1}
 &(5.19) \cr}
$$
where we have introduced the notation $y\equiv -\partial_+ \phi^{-1}_0$. In
terms of
these
quantities,  Eq. (5.13) yields the results
$$
\eqalignno{
\delta x^+&=-\gamma^0_{-1}&(5.20a)\cr
\delta \phi^{-1}_0 &=-\gamma^{-1}_0.&(5.20b)\cr
\delta \phi^1_{-2} &=-\gamma^1_{-2}-2\phi^1_{-2}\partial_+\gamma^0_{-1}+
                    \partial_+\phi^1_{-2}\gamma^0_{-1} &(5.20c) \cr
 \delta\phi^2_{-3} &=-\gamma^2_{-3}+\gamma^0_{-1}\partial_+\phi^2_{-3}
-3\partial_+\gamma^0_{-1}\phi^2_{-3} +\gamma^1_{-2}\partial_+\phi^1_{-2}
-\partial_+\gamma^1_{-2}\phi^1_{-2} &(5.20d) \cr
& \ \vdots  \cr}
$$
Note that only fields and parameters corresponding to the left edge, i.e.
$\phi^\ell_{-\ell-1}$ occur
in these results. Moreover, Eq. (5.20b) is precisely the $w_{1+\infty}$
transformation
rule of
Eq.\ (4.12), after identifying $\phi$ and $k_\ell$ with  $(-\phi^{-1}_0)$
and
$\beta^\ell_{-\ell-1}$, respectively.

It may be verified that the action
${\cal L}_0 = \ft12\partial_+\phi^{-1}_0\partial_-\phi^{-1}_0$
transforms by a total derivative under the full set of
$w^\uparrow_{1+\infty}$
transformations
$\delta \phi^{-1}_0 = -\sum_{\ell=-1}^\infty k^\ell y^{\ell+1}$.
 Free second-order
scalar
actions involving the higher left-edge Goldstone fields
$\phi^\ell_{-\ell-1}$ cannot
be made because
they do not have Lorentz weight zero, since the Lorentz weight of
$\phi^\ell_{-\ell-1}$
is
$-(\ell+1)$. One may, however, couple the higher Goldstone fields to
currents built
from
$\phi^{-1}_0$.  One way to this would be to construct a composite
$w_{1+\infty}$
connection
$A^\ell_{-\ell-1}$ built out  of the ``edge scalars'', and to use it in
 the
$w_\infty$ gravity action of Ref.~[11], thus obtaining the global
$w_{1+\infty}$
symmetric
Lagrangian  $$
{\cal L}=\ft12\partial_+\phi^{-1}_0\partial_-\phi^{-1}_0
-\sum_{\ell=0}^\infty{1\over \ell+2}
       A_{(-)\,-\ell-1}^\ell(\phi)(-\partial_+\phi^{-1}_0)^{\ell+2}.
\eqno(5.21)
$$
The construction of the composite connection requires further work. The idea is
essentially as outlined above, though, the construction may have to be
 based on a
different coset
space than the one considered here [15].

\bigskip
\noindent{\bf 6. Quantum Realizations of $w_\infty$}
\bigskip

The realization of $w_{1+\infty}$ in terms of the currents
$$
     v^\ell(z) = {1\over \ell+2} (\p_+ \phi)^{\ell+2}  \eqno(6.1)
$$
is necessarily a classical one, since in commuting two of these currents
which
involves an operator
product expansion, there will be terms coming from multiple Wick
contractions of the
 basic
building blocks $\p_+\phi$, giving rise to terms that violate the closure
of the
algebra. The single
contraction, on the other hand, corresponds to evaluating the Poisson
bracket of
the two currents,
evidently gives a closed algebra. If one modifies the currents (6.1) as
$V_\ell(z)=\sum c^\ell_{mnp}({\sqrt \hbar})^m (\p^n \phi)^p$ where
 $c^\ell_{mnp}$
are a set of
constant coefficients, then the algebra will of course close on these
currents.
(On
dimensional grounds $(n+1)p+m=2\ell+2$). Although this closure may in some
sense be
considered as
trivial, in fact there is a way of choosing the coefficients (which
amounts to using
a certain basis
for the algebra) in such a way that the currents $V^\ell(z)$ have a
definite
transformation property
under an $SL(2,R)$ and moreover they are quasi-primary fields with
respect to
a natural Virasoro
subalgebra. The resulting algebra is the $W_{1+\infty}$ algebra. At the
level
of quantum
field theories this phenomenon amounts to the deformation of the classical
$w_{1+\infty}$ symmetry to
a quantum $W_{1+\infty}$ symmetry [12]. The former can be obtained in the
$\hbar \rightarrow 0$ limit
of the latter one. This phenomenon has been found in the study of quantum
$w_\infty$ gravity as well
as the study of a quantum mechanical system on a circle [13].

Turning our attention to the issue of constructing a quantum realization of
 $w_{1+\infty}$, one
possibility is to construct the $w_{1+\infty}$ currents in terms of b-c
ghost systems.
As a
first step, one constructs the BRST charge [33]
$$
  Q_{\rm gh}=\oint f^{ij}(\p,\p)\big(-\ft12 c_ic_jb_{i+j}\big),  \eqno(6.2)
$$
where $b_i(z)$ and $c_i(z)$ are anticommuting ghosts satisfying
$$
b_i(z)c_j(w)\sim {\delta_{ij}\over z-w},\eqno(6.3)
$$
or equivalently, the anticommutator $\{b_i, c_j\}=\delta_{ij}$ and
$f^{ij}(m,n)=(j+1)m-(i+1)n$ are
the structure constants of $w_{1+\infty}$. The notation  $f^{ij}(\p,\p)$
indicates that the
Fourier-mode index $m$ is replaced by a partial derivative that acts on $c_i$
only, and the index $n$
is replaced by a partial derivative that acts on $c_j$ only. From
$Q_{\rm gh}$,
we can derive the
ghostly quantum realization $v^i_{\rm gh}$ of $w_{1+\infty}$
$$
   v^i_{\rm gh}=\{Q_{\rm gh}, b_i\}, \eqno(6.4)
$$
This formula yields [33]
$$
v^i_{\rm gh}(z)=\sum_{j\ge0}\Big( (i+j+2)\partial c_j\,
b_{i+j} +(j+1) c_j\,
\partial b_{i+j}\Big),\eqno(6.5)
$$
which indeed generate the $w_\infty$ algebra:
$$
v^i_{\rm gh}(z)v^j_{\rm gh}(w)\sim {i+j+2\over (z-w)^2}v^{i+j}_{\rm gh}(w)
+
{i+1\over z-w}\partial
v^{i+j}_{\rm gh}(w) +\delta^{i0}\delta^{j0} {c/2\over (z-w)^4}.\eqno(6.6)
$$
The operator terms on the right-hand side come from single contractions.
The central term in the spin-2 sector (the only one that occurs in the
$w_\infty$ algebra) has a central charge that is formally divergent.
After zeta-function
regularization, one  finds $c=2$ [34,33].

Using the structure constants of the topological algebra
$w_\infty^{\rm top}$ given
in (3.2), by means
of the method outlined above, one can also construct its ghostly quantum
 realization
as follows [26]
$$
\eqalign{
         {\hat v}_{\rm gh}^i(z)&=(i+j+2)\partial c_j
b_{i+j}+(j+1)c_j\partial
b_{i+j}-(i+j+2)\partial \gamma_j \beta_{i+j}-(j+1)\gamma_j\partial
\beta_{i+j},
\cr
G^i_{\rm gh}(w)&=(i+j+2)\partial c_j\beta_{i+j}+(j+1)c_j
\partial\beta_{i+j},
\cr}\eqno(6.7)
$$
where $\beta,\ \gamma$ are the commuting ghost fields, and summation over
$j$ is
understood. A
detailed description of topological $w_{1+\infty}$ is given in Ref.~[26].

Finally, let us mention a quantum realization of the loop algebra of
${\rm SDiff}(R^2)$
due to Witten
[23], which has emerged in the study of string theory with two dimensional
target
spacetime. The
theory is characterized by the stress tensor
$$
T_{zz}= -\ft12 (\p X)^2 -\ft12 (\p\phi)^2 +{\sqrt 2}\phi-2b\p c+c\p b
\eqno(6.8)
$$
where $X$ is the matter field, $\phi$ is the Liouville field and $b,\ c$
are the ghost
fields, all
of which obey the usual OPE rules. A quantum realization of of the loop
algebra of
${\rm SDiff}(R^2)$  is given by [23]
$$
v^s_n(z)=\Big(e^{-i{\sqrt 2}X}\Big)^{s-n} e^{{is\over \sqrt 2}X}e^{{\sqrt 2}
(1-s)\phi}
\eqno(6.9)
$$
Acting on a polynomials in [23]
$$
\eqalign{
x &=\Big(cb+{i\over \sqrt 2}(\p X-i\p \phi)\Big) e^{{i\over \sqrt 2}(X+i\phi)}
 \cr
y &= \Big(cb-{i\over \sqrt 2}(\p X+i\p \phi)\Big) e^{-{i\over \sqrt 2}
(X-i\phi)},  \cr}
\eqno(6.10)
$$
the generators defined in (6.9) obey the algebra (3.6). In particular
$v^{1/2}_{1/2}$ acts like ${\p \over \p y}$ and  $v^{1/2}_{-1/2}$ acts
like
${\p \over \p x}$, and hence commute, on the plane. However, as operators
they do not
commute,
yielding a central extension term [23].

\bigskip
\noindent{\bf 7. Comments}
\bigskip
The subject of $w_\infty$ algebras is a rapidly growing one. Here we have
attempted to
give some of
the salient features of this subject, and necessarily have omitted a number
of topics,
 some of which
are briefly mentioned in the introduction. Many of these topics are relevant
in one
way or another
to the question of how to construct new string theories with higher
world-sheet, and
possibly higher spacetime symmetries.

The $w_\infty$ algebra is a certain $N\rightarrow \infty$ limit
 [35] of the $W_N$ algebra [36]. Some work has already been done on string
theories
based on $W_N$
symmetry [37,38,39]. A typical feature which has arisen is that
although one expects a higher slope and therefore  higher spin massless
fields in the
spectrum
[40], it turns out that due to the necessary presence of a background
charge at least
for one of
the scalars in the theory, the slope is pushed back to a value in such a way
that there
are no
massless higher spin fields after all [38,39]. At the end, one finds the
spectrum of the
usual string
and new massive trajectories. In this author's opinion this state of affairs is
somewhat
disappointing, because the symmetry enlargement one intuitively expects
is buried in the
complexities of the massive trajectories, if at all there. Of course,
there may be new
field
theoretic realizations of $W$-algebras still to be discovered where the
 situation may
be dramatically
different.

A string theory based on $w_\infty$, or
its quantum deformation $W_\infty$ has not been considered so far.
It would be very
useful to
accumulate field theoretic realizations of these symmetries which might
play a role
in constructing
a sensible $w_\infty$ string theory. Such a theory may as well look like
a topological
field theory,
since there would be infinitely many physical state conditions to satisfy.
 We would
like to
speculate, however, that there may exist a $w_\infty$  string theory where
{\it only} a
tower of
higher spin {\it massless} fields in {\it target} spacetime would arise in
the spectrum,
corresponding to the infinitely many  higher spin world-sheet symmetries,
and that the
usual string
theory with its massive states may arise as a result of some sort of
``spontaneous''
symmetry
breaking. Clearly a lot remains to be done, and there could be some
surprises ahead
in the search for
a string theory (or possibly a theory of higher extended objects, such
as supermembranes)  with higher
symmetries than those which have  been realized so far.

\bigskip

\centerline{ACKNOWLEDGEMENTS}

\bigskip

I would like to thank Eric Bergshoeff, Steve Fulling, Chris Pope and Kelly
Stelle for
discussions. I also would like to thank Professor Abdus Salam, the
International Atomic Energy Agency and UNESCO for hospitality at the
International Center for Theoretical Physics where lectures on the topics
of this
review article were
delivered. This work is partially supported by NSF grant PHY-9106593.

\vfil\eject

\centerline{\bf REFERENCES}
\bigskip
\frenchspacing

\item{1.} V.I. Arnold, {\it Mathematical Methods of Classical Mechanics},
Springer
           Verlag,1978; Ann. Inst. Fourier {\bf 16} (1966) 319.
\item{}    J. Moser, Siam Review {\bf 28} (1986) 459.
\item{2.} J. Hoppe, MIT Ph. D. Thesis, 1982 and in {\it Proc. Int. Workshop
            on Constraints Theory and Relativistic Dynamics},
            Eds. G. Longhi and L. Lusanna (World Scientific, 1987).
\item{3.} E. Bergshoeff, E. Sezgin and  P.K. Townsend, Ann. Phys. {\bf 199}
 (1990) 340.
\item{4.}  B. de Wit, J. Hoppe and H. Nicolai,Nucl. Phys. {\bf B305} (1988)
545.
\item{5.} E.G. Floratos and J. Illiopoulos, Phys. Lett. {\bf 201B} (1988)
237.
\item{6.} I. Antoniadis, P. Ditsas, E.G. Floratos and J. Illipoulos,
 Nucl. Phys.
         {\bf B300} (1988) 549.
\item{7.}I. Bakas, Phys. Lett. {\bf 228B} (1989) 57.
\item{8.} C.N. Pope, L.J. Romans and X. Shen, Phys. Lett. {\bf 236B} (1990)
173.
\item{9.} I. Bars, C.N. Pope and E. Sezgin, Phys. Lett. {\bf 210B} (1988)
85.
\item{10.} C.N. Pope, L.J. Romans and X. Shen, Nucl. Phys. {\bf B339} (1990)
191.
\item{11.} E. Bergshoeff, C.N. Pope, L.J. Romans, E. Sezgin, X. Shen and
           K.S. Stelle,\nl{\sl Phys. Lett.} {\bf 243B} (1990) 350.
\item{12.} E. Bergshoeff, P.S. Howe, C.N. Pope, E. Sezgin, X. Shen and K.S.
            Stelle,\nl{\sl Nucl. Phys.} {\bf B363} (1991) 163.
\item{13.} R. Floeranini, R. Percacci and E. Sezgin, Phys. Lett. {\bf 261B}
(1991) 51.
\item{14.} C.M. Hull, Phys. Lett. {\bf 269B} (1991) 257; {\it Classical and
 quantum
W-gravity},
            preprint, QMW/PH/92/1 (January 1992).
\item{15.} E. Sezgin and K.S. Stelle, {\it Nonlinear realizations of
$w_\infty$},
preprint, CTP
            TAMU-26/91, Imperial/TP/90-91/21.
 \item{16.} Q.H. Park, Phys. Lett. {\bf 236B} (1990) 429; Phys. Lett.
{\bf 238B} (1990) 287;Phys.
           Lett. {\bf 257B} (1991)105.
\item{} K. Yamagishi and G.F. Chapline, Class. \& Quantum Grav. {\bf 8}
(1991) 427;
Phys. Lett. {\bf
         259B} (1991) 463.
\item{17.} H. Ooguri and C. Vafa, Mod. Phys. Lett. {\bf A5}(1990) 1389;
Nucl. Phys.
{\bf B361}
           (1991) 469; Nucl. Phys. {\bf B367} (1991) 83.
\item{18.} E. Bergshoeff, M. Blencowe and K.S. Stelle, Commun. Math. Phys.
{\bf 128}
(1990) 213.
\item{19.} T.A. Arakelyan and G.K. Savvidy, Phys. Lett. {\bf 214B} (1988)
350.
\item{20.} M.P. Blencowe, Class. \& Quantum Grav. {\bf 6} (1989) 443.
\item{21.} E.S. Fradkin and M. Vasiliev, Ann. Phys. {\bf 177} (1987) 63.
\item{22.} C.N. Pope and X. Shen, Phys. Lett. {\bf 236B} (1990) 21.
\item{23.} E. Witten, {\it Ground Ring of Two-dimensional String Theory},
preprint,
IASSNS-HEP-91/51
            (August 1991).
\item{24.} E. Sezgin, in {\it Strings '89}, Eds. R. Arnowitt, R. Bryan,
             M.J. Duff, D. Nanopoulos and C.N. Pope (World Scientific, 1990).
\item{25.} E. Sezgin, {\it Aspects of $W_\infty$ symmetry}, preprint,
IC/91/206,
CTP TAMU-9/91.
\item{26.} C.N. Pope, L.J. Romans, E. Sezgin and X. Shen, Phys. Lett.
{\bf 256B} (1991)
191.
\item{27.} E. Witten, Comm. Math. Phys. {\bf 117} (1988) 353.
\item{28.} S. Odake and T. Sano, Phys. Lett. {\bf 258B} (1991) 369.
\item{29.} I. Bakas and E. Kiritsis, in {\it Common trends in mathematics
 and quantum
field
         theories}, Eds. T. Eguchi, T. Inami and T. Miwa (Prog. Theor.
Phys.
Supplement, 1990).
\item{30.} K. Schoutens, A. Sevrin and P. van Nieuwenhuizen, Phys. Lett.
{\bf 243B}
(1990) 245.
\item{31.} C.M. Hull, Phys. Lett. {\bf 240B} (1989) 110.
\item{32.} E. Bergshoeff and M. de Roo, {\it $N=2\ w_\infty$ supergravity},
preprint,
 UG-62/91
          (September 1991).
\item{33.}C.N. Pope, L.J. Romans and X. Shen, Phys. Lett. {\bf 242B} (1990)
401.
\item{34.} K. Yamagishi, Phys. Lett. {\bf 266B} (1991) 370.
\item{35.} I. Bakas, Commun. Math. Phys. {\bf 134} (1990) 487.
\item{36.}A.B. Zamolodchikov, Teo. Mat. Fiz. {\bf 65} (1985) 347;
\item{} V.A. Fateev and S. Lykyanov, Int. J. Mod. Phys. {\bf A3} (1988)
507.
\item{37.} S.R. Das, A. Dhar and S.K. Rama, {\it Physical properties of W
gravities
 and W strings},
            preprint, TIFR/TH/91-11.
\item{38.} C.N. Pope, L.J. Romans, E. Sezgin and K.S. Stelle, Phys. Lett.
{\bf 274B}
(1992) 298.
\item{39.}    H. Lu, C.N. Pope, S. Schrans and K.W. Xu, {\it The complete
spectrum of
the $W_N$
           string}, preprint, CTP TAMU -5/92, KUL-TF-92/1 (January 1992).
\item{40.} A. Bilal and J.L. Gervais, {\sl Nucl.\ Phys.}\ {\bf B326} (1989)
222.

\end